\documentclass{article}
\usepackage{spconf,amsmath,graphicx}
\usepackage{amsfonts, hyperref}

\title{A Cross-Verification Approach for Protecting World Leaders from Fake and Tampered Audio}
\name{Mengyi Shan \qquad TJ Tsai}
\address{Harvey Mudd College, Claremont, CA USA}
\begin{document}
\ninept
\maketitle

\begin{abstract}
This paper tackles the problem of verifying the authenticity of speech recordings from world leaders.  Whereas previous work on detecting deep fake or tampered audio focus on scrutinizing an audio recording in isolation, we instead reframe the problem and focus on cross-verifying a questionable recording against trusted references.  We present a method for cross-verifying a speech recording against a reference that consists of two steps: aligning the two recordings and then classifying each query frame as matching or non-matching.  We propose a subsequence alignment method based on the Needleman-Wunsch algorithm and show that it significantly outperforms dynamic time warping in handling common tampering operations.  We also explore several binary classification models based on LSTM and Transformer architectures to verify content at the frame level.  Through extensive experiments on tampered speech recordings of Donald Trump, we show that our system can reliably detect audio tampering operations of different types and durations. Our best model achieves 99.7\% accuracy for the alignment task at an error tolerance of 50 ms and a 0.43\% equal error rate in classifying audio frames as matching or non-matching.
\end{abstract}
\section{Introduction}
\label{sec:intro}
This paper studies the problem of verifying the authenticity of speech recordings from world leaders.  Recent developments in technology are making it possible to generate or manipulate audiovisual information in a way that looks and sounds very realistic \cite{Kietzmann_2019_deepfake}.  This has sparked concern over malicious uses, particularly in misrepresenting high-profile political leaders \cite{Agarwal_2019_CVPR_Workshops}.

One way in which a world leader may be misrepresented is through a deep fake impersonation.  Existing methods can modify videos in a photorealistic manner to lip-sync to unrelated audio recordings \cite{Suwajanakorn_2017_obama} or allow a source actor to control the facial expressions and head movements of a person in a video \cite{Thies_2016_face2face}\cite{Kim_2018_deepvideo}.  In a similar vein, advances in speech synthesis have enabled systems to imitate the characteristics of a person's voice with very limited training data \cite{Arik_2018_voiceclone}\cite{Jia_2018_synthesis}.  There has been a flurry of recent works on detecting fake audiovisual data, many of which focus on finding inconsistencies between or within the audio and video modalities \cite{Chintha_2020_RCStructure}\cite{Agarwal_2020_deepfake}\cite{Agarwal_2019_CVPR_Workshops}\cite{Li_2018_eyeblinking}.  The ASVspoof \cite{Todisco_2019_ASV} competition has also spurred research on detecting various forms of spoofed audio, including synthetic, converted, and replayed audio.

Another way in which a world leader may be misrepresented is through audio tampering. Audio tampering involves modifying a genuine audio recording through insertion, deletion, or replacement.  The availability of audio editing software (e.g.~Audacity, Adobe Audition) and other tools makes it easy to modify audio recordings through cutting, concatenating, and combining audio segments (e.g. see \cite{Jin_2017_Voco} for a sophisticated approach).  The field of audio forensics provides a rich set of methods for detecting such tampering operations.  These methods include looking for inconsistencies in the embedded electrical network frequency \cite{Reis_2017_ESPRIT}\cite{Wang2018DigitalAT}, acoustic environment signature \cite{patole2017reverberation}\cite{Bhangale_2019_Reverberation}, microphone signature \cite{Cuccovillo_2013_microphone}\cite{Moon_2014_reducenoise}, or compression quantization characteristic \cite{Yang_2012_frame}\cite{Tao_2016_quantization}.  See \cite{Zakariah2016DigitalMA} for a recent survey.

The main difference between our proposed approach and previous work is in our framing of the problem.  Rather than scrutinizing an audio recording in isolation to determine if it is fake or tampered, we instead focus on positively verifying speech recordings against existing sources.  This is \textit{not} a solution to the general problem of fake and tampered audio, but it \textit{is} a viable solution to the specific problem of protecting world leaders, whose public speeches and announcements are recorded by multiple news agencies.

This approach has several significant benefits. First, the efficacy of such a system does not depend in any way on the quality of a deep fake recording or how seamlessly a tampering operation is done.  By focusing entirely on cross-verifying with other sources (rather than scrutinizing the artifact itself), fraudulent recordings can be exposed even if the fake or tampered audio is so convincing as to be indistinguishable by both human and machine.  In this way, the methodology would provide a permanent solution, rather than a temporary fix that depends on the state of deep fake or audiovisual editing technology.  Second, the proposed approach would provide a way to positively verify the authenticity of real recordings.  Beyond simply detecting fake or tampered content, our approach provides a way to confidently conclude that a recording is genuine.

This paper has two main contributions.  First, we introduce an approach to the problem of protecting world leaders from fake or tampered audio that focuses on cross-verification against a database of trusted reference recordings.  Second, we propose a two-stage method for performing this cross-verification by first aligning the query and reference recording, and then classifying every frame in the query as matching or non-matching.  We introduce an alignment algorithm that is a variant of Needleman-Wunsch \cite{needleman1970general} and show that it significantly outperforms dynamic time warping, and we experiment with several binary classification models based on LSTM and Transformer architectures.  Through detailed experiments on a range of conditions and tampering operations, we show that the proposed approach performs robustly.\footnote{Code for reproducing the results can be found at \url{https://github.com/HMC-MIR/TamperingDetection}.}

\section{System Description}
In this section, we describe our proposed solution to verify speech recordings from world leaders.  In the next four subsections, we will describe the high-level approach (Section \ref{subsec:approach}) and details for the cross-verification subsystem (Sections \ref{subsec:features}, \ref{subsec:system_alignment}, \ref{subsec:system_attribution}).

\subsection{High-Level Approach}
\label{subsec:approach}

\begin{figure}
    \centering
    \includegraphics[width=.8\columnwidth]{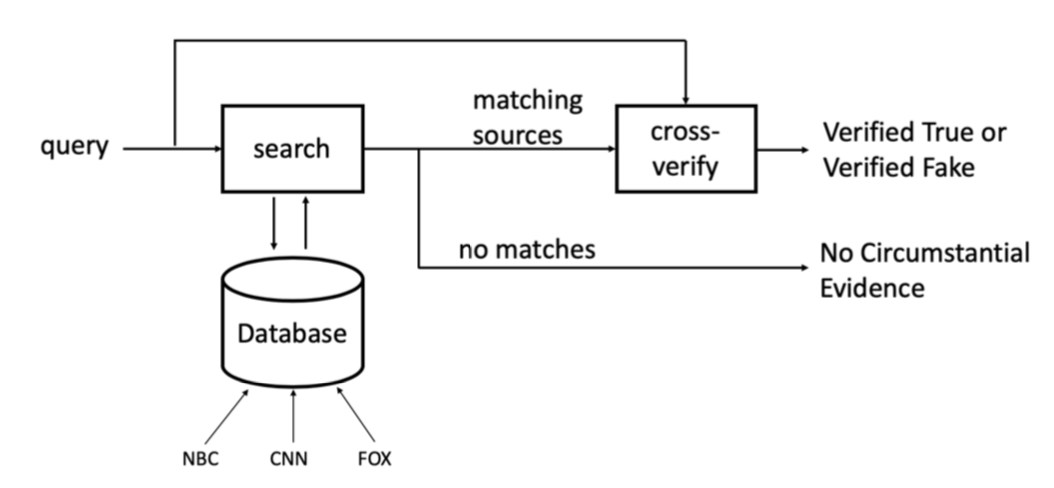}
    \caption{Overview of the whole system.  The goal is to verify the authenticity of a speech recording given by a world leader.}
    \label{fig:system}
\end{figure}

Figure \ref{fig:system} shows a block diagram of the high-level approach.  There are three main components: (a) a database containing raw, unedited recordings from trusted and reliable sources such as major news agencies, (b) a search sub-system that, given an audio query, finds a match in the database, and (c) a cross-verification sub-system that compares the audio query with a matching reference recording and verifies the legitimacy of the content at a frame level.  In this work, we focus exclusively on the audio modality, though other modalities like video or metadata (e.g. the approximate date of the recording in question) could be incorporated into the same framework.

To gain more intuition into how this system performs, it is helpful to consider how the system would respond in three different cases of interest.  Case 1: If the query is a truthful recording, the system will find a match in the database, and the cross-verification system will confirm that the query is truthful.  Case 2: If the query is a deep fake, the search will fail to find any matches in the database since the message is completely made-up.  In this case, the system concludes that there is no circumstantial evidence to corroborate the recording.  The recording could in fact be real, but the system can conclude that the statement was not publicly made to the press as an official announcement.  Case 3: If the query is a real recording that has been tampered, the search mechanism will likely still find a match in the database (unless the tampering is so severe that there is little resemblance to the original), and the cross-verification module would detect the portions of the query that have been tampered.

The search problem can be considered solved for all practical purposes.  There are many effective existing approaches for performing audio-audio search (e.g.~\cite{baluja2008waveprint}\cite{Wang03anindustrial-strength}\cite{Jaap_2003_fingerprint}), and many commercial systems are widely used for copyright detection and song identification \cite{gfeller2017now}\cite{Wang03anindustrial-strength}.  Because our scenario of interest consists of clean audio recordings, we can expect these systems to perform robustly as long as there are at least a few seconds of matching audio.

\begin{figure}
	\centering
	\includegraphics[width=\columnwidth]{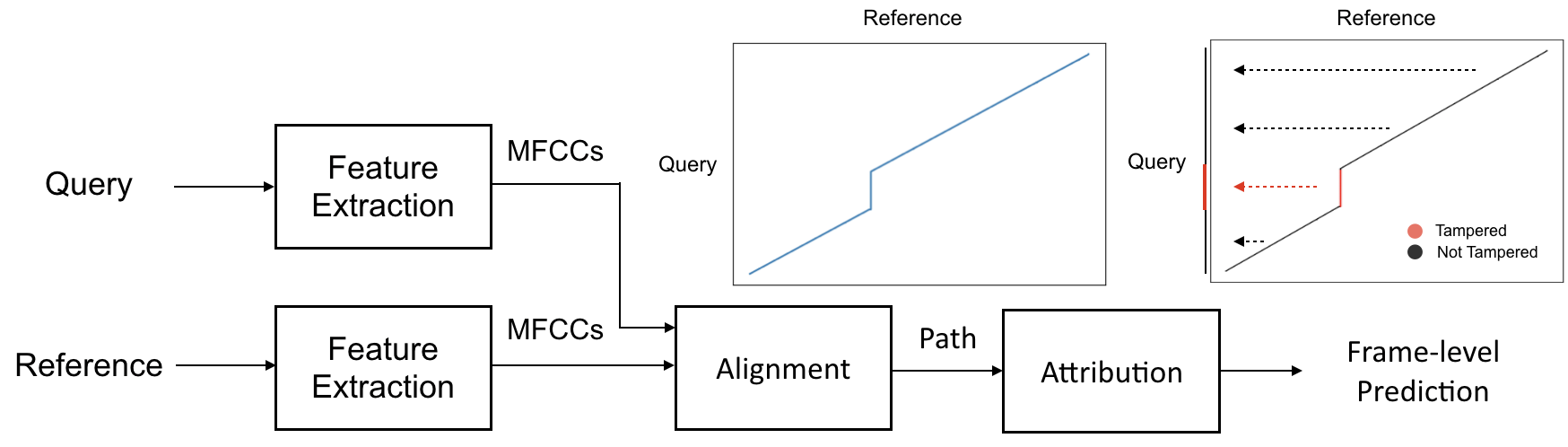}
	\caption{Overview of the cross-verification approach.}
	\label{fig:cross-verification}
\end{figure}

The main unsolved technical challenge in Figure \ref{fig:system}, therefore, is the cross-verification sub-system.  Figure \ref{fig:cross-verification} shows the three steps in our cross-verification approach.  These three steps will be described in more detail in the next three subsections.

\subsection{Feature Extraction}
\label{subsec:features}

The first step in cross-verification is to extract features from both the query and reference audio recordings.  Note that the reference audio is the matching recording found by the audio search sub-system.  This reference recording contains the (possibly tampered) audio query somewhere within it.  We compute Mel-frequency cepstral coefficients (MFCCs) with standard settings for speech audio: 13-dimensional MFCCs with delta and delta-delta features (total 39 dimensions), 25 ms analysis frames, and 10 ms hop size.  At the end of this first step, we have two sequences of features $x_1,\dots, x_N$ and $y_1,\dots, y_M$ where $x_i \in \mathbb{R}^{39}$ and $y_i \in \mathbb{R}^{39}$.

\subsection{Alignment}
\label{subsec:system_alignment}

The second step in cross-verification is to align the query and reference features.  We initially tried using dynamic time warping (DTW) with various hyperparameter settings but found that the performance was unacceptably poor.  Below, we describe an alternative alignment algorithm that we found to be effective.  In Section \ref{subsec:alignment}, we will provide both an empirical evaluation and an intuitive explanation for DTW's poor performance.

Our alignment approach is a modified subsequence Needleman-Wunsch time warping  (SNWTW) algorithm \cite{needleman1970general}.  We first compute a pairwise cost matrix $C \in \mathbb{R}^{N \times M}$ between the two feature sequences using a Euclidean distance measure.  So, $C[i,j]$ indicates the dissimilarity between the query feature $x_i$ and the reference feature $y_j$.  Next, we compute a cumulative cost matrix $D \in \mathbb{R}^{N \times M}$ (and corresponding backtrace matrix $B$) using the following dynamic programming rules:
\small
\begin{equation*}
  D[i,j] =
    \begin{cases}
      C[i,j] & \text{$i=0$}\\
      \alpha + D[i-1, j] & \text{$i>0, j=0$}\\
      \min (\gamma + D[i, j-1], \alpha + D[i-1, j],\\ 
            \quad \quad C[i, j] + D[i-1, j-1])  & i>0, j>0
    \end{cases}       
\end{equation*}
\normalsize
Note that there are three types of transitions: (i) moving up, which is skipping a frame in the query dimension, (ii) moving right, which is skipping a frame in the reference dimension, and (iii) moving diagonally, which corresponds to a match.  $\alpha$ and $\gamma$ are penalties for skipping in the query and reference dimensions, respectively.  The initialization of $D$ allows the subsequence path to start at any point along the reference sequence without penalty.   As we compute each element $D[i, j]$, we also update $B[i, j]$ to record the optimal transition type at each location.  Once $D$ is computed, we identify the lowest cost element in the last row of $D$ and follow the back pointers to determine each step of the optimal subsequence path.

Our proposed method has two hyperparameters: $\gamma$ and $\alpha$.  For the horizontal skip penalty $\gamma$ (skipping in the reference dimension), we found that the optimal setting is a small, positive number.  We want the cost to be small in order to allow skipping in the reference dimension when there is a deletion edit in the query.  If we set the cost to zero, however, it allows for arbitrarily long skips, which has undesirable behavior.  In our experiments, we found that the value of $\gamma$ is relatively insensitive, having approximately the same optimal performance across almost two orders of magnitude (0.1 to 5).  We set $\gamma = 1$ for all reported results.  For the vertical skip penalty $\alpha$ (skipping in the query dimension), we found that the optimal value of $\alpha$ strongly depends on the values in the cost matrix.  If $\alpha$ is too large, the algorithm cannot follow vertical alignment paths corresponding to insertions.  If $\alpha$ is too small, the optimal subsequence path consists of simply skipping across the entire cost matrix.  Ideally, we would like the penalty to be greater than the cost between two matching elements but smaller than the cost between non-matching elements.  Accordingly, we set $\alpha$ in a data-dependent way: we calculate the minimum cost in each row of the cost matrix $C$, and then set $\alpha$ to two times the median of these minima.  Thus, the vertical penalty is set to twice the estimated cost between two matching frames, where we estimate the cost between matching frames as the row minima.  

At the end of the second step, we have a predicted alignment path between the query and reference.  Figure \ref{fig:cross-verification} shows an alignment path with a query modified by insertion.

\subsection{Attribution}
\label{subsec:system_attribution}

The third step in cross-verification is to classify each audio frame in the query as matching or not matching.  We refer to this as the attribution task.  Frames that are matching are considered to be positively verified against the reference.  This is done in three sub-steps.

The first sub-step is to convert the predicted alignment path into a sequence of features.  For every element in the alignment path, we identify the optimal back pointer (i.e.~up, right, or diagonal) from the backtrace matrix $B$.  This is encoded as a 3-dimensional one-hot encoded vector.  We multiply this one-hot encoded vector with the corresponding pairwise cost at that location in the cost matrix $C$.  Thus, the feature representation encodes both the optimal transition type as well as the pairwise dissimilarity.  The alignment path is represented as a sequence of 3-dimensional features.

The second sub-step is to process the sequence of path features with an LSTM or Transformer model.  For each element of the alignment path, the model returns a prediction of how likely the path features correspond to a match or non-match.  We experiment with both LSTM and bidirectional LSTM models with 1, 2, and 3 layers, all with a hidden layer size of 32 and an output classification layer.  For the Transformer, we consider Transformer encoder models with 1, 2, and 3 layers, all with three attention heads \footnote{Since the number of attention heads must divide the embedding dimension size (3) evenly, the number of attention heads can only be 3 or 1.} and a final classification layer.  Note that, since we want context information to come from both directions, we use Transformer encoder layers rather than Transformer decoder layers.

The third sub-step is to impute the path element predictions to their corresponding query audio frames.  Each frame in the audio query can be uniquely associated with an element on the predicted alignment path.  When deletions occur and there is a long horizontal segment in the alignment path, the leftmost element of the horizontal segment should be used for the corresponding alignment.  This provides the final output of the attribution model: a score for each query audio frame that indicates if it matches the reference recording.

At the end of the third step, we have a prediction of which query frames match the reference, and which frames in the reference they correspond to.  These constitute the final outputs of our system.

\section{Experimental Setup}

The dataset comes from recordings of US President Donald Trump. We collected 50 speech recordings of Trump from 2017 to 2018, all taken from the official white house Youtube playlist. Each of these recordings is single-speaker and is 1-3 minutes long.  Two of the 50 recordings have 20 seconds of background music accompanying part of the speech. These recordings are included included in the dataset.

We generate audio queries from the raw data in the following manner.  We randomly select ten 10-second segments from each recording, and then generate four different versions of each segment by applying various tampering operations.  The first version has no tampering: it is simply an unedited version at a lower bitrate quality (160kbps) than the reference (320 kbps).  The second version has an insertion: we randomly select an $L$ second fragment from a different audio recording and insert the fragment into the 10-second segment at a random location.  The third version has a deletion: we randomly select an $L$ second interval from the segment and delete it.  The fourth version has a replacement: we randomly select an $L$ second fragment from a different audio recording and replace a randomly selected $L$ second fragment in the 10-second segment.  For a given value of $L$, this process will generate a total of $4 \times 50 \times 10 = 2000$ queries.  We divide the queries by speech recording and use 50\% for training, 20\% for validation, and 30\% for testing.

Preparing the training data for the attribution models requires propagating the query \& reference frame labels (i.e. match vs non-match) to the predicted alignment path elements.  For example, if an alignment path element corresponds to a query or reference frame that is inserted, deleted, or replaced, the path element will be labeled as a non-match.  In this way, we can propagate ground truth labels to predicted alignment path elements in a deterministic manner.

\begin{table}
	\centering
	\begin{tabular}{|l|c|c|c|c|c|}
		\hline
		System              & \multicolumn{5}{c|}{Tolerance}        \\
		& 20ms     & 50ms     & 100ms   & 200ms    & 500ms    \\ \hline
		DTW (1,1,1) & 4.17\% & 3.70\% & 3.69\% & 3.67\% & 3.63\% \\ \hline
		DTW (0,1,1) & 3.06\% & 2.90\% & 2.85\% & 2.81\% & 2.79\% \\ \hline
		DTW (1,2,1) & 9.97\% & 9.70\% & 9.68\% & 9.64\% & 9.63\% \\ \hline
		SNWTW         & 0.77\% & 0.29\% & 0.25\% & 0.17\% & 0.14\% \\ \hline
	\end{tabular}
	\caption{Performance of DTW and the proposed SNWTW on the alignment task.  Numbers indicate error rates at a specified tolerance.}
	\label{tab:alignment-dtw}
\end{table}

We evaluate the alignment and attribution tasks separately.  For the alignment task, we evaluate performance by (a) calculating the alignment error for each query frame that has a true match in the reference, (b) determining the percentage of query frames whose alignment error is below a fixed error tolerance, and (c) sweeping across a range of different error tolerances to characterize the tradeoff between error rate and error tolerance.  Note that the alignment task only evaluates query frames that have a (ground truth) match in the reference; it ignores frames that are the result of tampering since they do not correspond to a frame in the reference.  For the attribution task, we evaluate the binary classification performance (match vs non-match) with a receiver operating characteristic (ROC) curve, which shows the tradeoff between true positives and  false positives for a range of threshold values.

Both the alignment and attribution tasks must be considered to characterize cross-verification performance.  Deletions cannot be explicitly handled by the attribution task (since all query frames do match the reference) but can be evaluated by the alignment task.  Thus, a system must perform well on both tasks in order to effectively solve the cross-verification problem.

\section{Results \& Analysis}
\label{sec:results}

\subsection{Alignment}
\label{subsec:alignment}

We compare the performance of four algorithms for the alignment task: the proposed SNWTW approach and subsequence DTW with three different hyperparameter settings.  All four systems have the same set of allowable transitions $\{(0,1),(1,0),(1,1)\}$, and the subsequence DTW variants use weightings of $\{1,1,1\}$, $\{0,1,1\}$, and $\{1,1,2\}$.  Table \ref{tab:alignment-dtw} compares the alignment error rate for all four systems across a range of tolerances.  These results are with 160 kbps audio queries and $L=2$ second edit durations.

\begin{figure}
	\centering
	\includegraphics[width=.8\columnwidth]{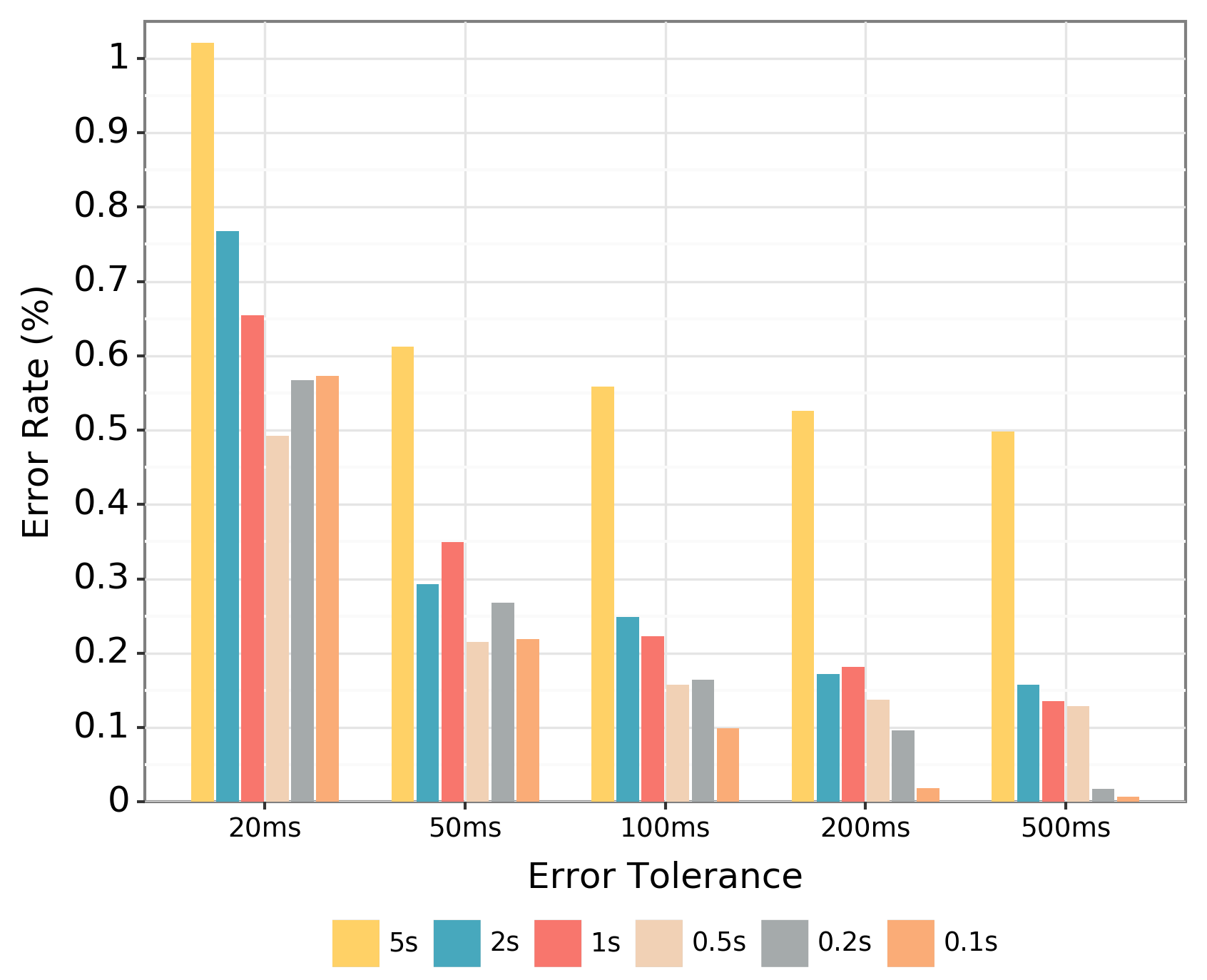}
	\caption{Performance of SNWTW on the alignment task with tampering operations of various duration.}
	\label{fig:alignment-result}
\end{figure}

There are two things to notice in Table \ref{tab:alignment-dtw}.  First, all subsequence DTW systems perform poorly on the alignment task, with the best error rate stabilizing around 2.8\% for the $\{0,1,1\}$ weighting.  The poor performance is because horizontal and vertical segments in the alignment path (corresponding to deletions and insertions) have extremely high costs.  Unlike SNWTW, vertical and horizontal segments in DTW are \textit{not} skips --- they are matches between completely unrelated audio frames, which results in very high costs that discourage DTW from correctly tracking insertions and deletions.  Second, SNWTW significantly outperforms the DTW systems and achieves an error rate of less than 0.3\% at 50 ms error tolerance.  It addresses the above issue with DTW by introducing the option to skip, where the skip penalty is independent of the (high) cost between non-matching audio frames.

We also compare the effect of different edit durations on the performance of SNWTW, as shown in Figure \ref{fig:alignment-result}.  Each group of bars corresponds to a fixed error tolerance, and each bar within the group shows the alignment error rate of SNWTW across six different edit durations.  Note that the y-axis is shown in units of percentage, so all systems have an error rate that is less than 1\%, and some have less than 0.2\% error rate.  We can see that small edit durations have smaller error rates, and that system performance is relatively stable across different edit durations except for the long 5 second edits.

We also ran additional experiments to determine the effect of query bitrate on alignment accuracy.  We found that bitrate has almost no impact on the results, even for low bitrates like 96kbps.

\subsection{Attribution}
\label{subsec:attribution}

We compare the performance of three different model architectures on the attribution task: LSTM, bidirectional LSTM, and Transformer.  For each architecture, we consider 1, 2, and 3 layer models.  All nine models are trained on the SNWTW alignment path data.

\begin{figure}
    \centering
    \includegraphics[width=.75\columnwidth]{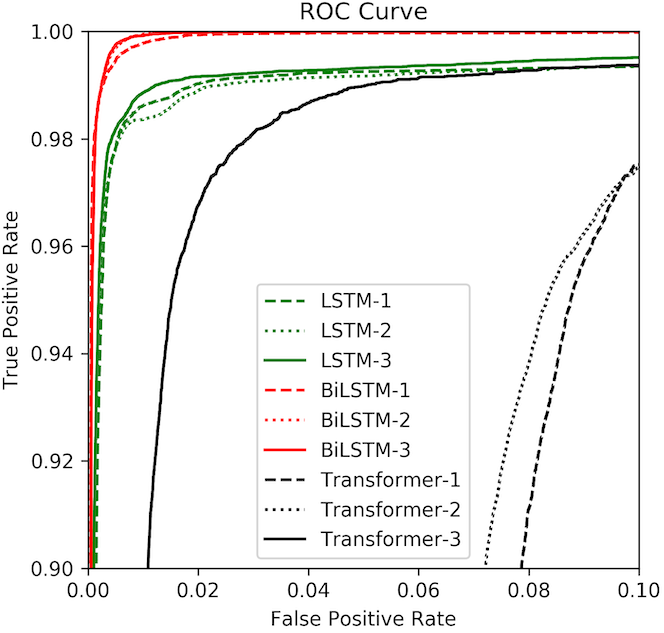}
    \caption{Performance of nine different attribution models.}
    \label{fig:attribution}
\end{figure}

Figure \ref{fig:attribution} compares the performance of all nine attribution models.  The results shown are for 160 kbps audio queries and $L=2$ second edits.  We can see that the bidirectional LSTM is significantly better than the other models, and that more layers lead to better performance, especially for the transformer model.  The equal error rates (EERs) for the 3-layer LSTM, BiLSTM, and Transformer models are $1.21\%$, $0.43\%$, and $1.82\%$, respectively.

We performed several additional experiments to determine the effect of edit duration on the best attribution model (3-layer BiLSTM) trained on $L=2$s data.  On the test benchmarks with $L$ equal to $5$s, $2$s, $1$s, $0.5$s, $0.2$s, and $0.1$s, the EER of the model was $1.20\%$, $0.43\%$, $0.53\%$, $1.28\%$, $5.54\%$, and $15.14\%$, respectively.  The performance on $L=2$ test data is best since the edit durations in training and test are matched.  As the length of the edit duration deviates more and more from $L=2$, we see the performance degrade.  It is worth pointing out that the performance is still relatively good for $L=1$s edits, and that it is unclear in a practical sense how much one could distort the meaning of an audio recording by making a single sub-second edit.  For long edits like $L=5$, the true detection rate is sufficiently high that many of the frames in the query would be flagged as not matching.

We also ran additional experiments to determine the effect of query bitrate on the attribution model.  We found (again) that bitrate has almost no impact on the results, even for low bitrates like 96kbps.

\section{Conclusion}
In this paper, we formulate the problem of verifying speech recordings of world leaders and propose a solution based on cross-verification with existing, trusted reference recordings.  We propose a cross-verification approach that consists of an alignment followed by an attribution model that labels individual audio frames as matching or non-matching.  We propose an alignment method based on the Needleman-Wunsch algorithm, and explore several classification models based on LSTM and Transformer architectures.  We provide extensive empirical validation and show strong performance across a range of realistic conditions.  

\section{Acknowledgments}
This material is based upon work supported by the National Science Foundation under Grant No.~1948531.  We gratefully acknowledge the support of NVIDIA Corporation with the donation of the GPU used for this research.


\vfill
\pagebreak

\bibliographystyle{IEEEbib}
\bibliography{tamperingDetection}

\end{document}